\documentclass[twocolumn]{article} 
\usepackage{graphicx,epsfig,latexsym}

\title{Universal properties in galaxies and cored DM profiles } 
\author{Paolo Salucci\\[1ex]
SISSA/ISAS, via Bonomea, 265, 34136 Trieste, Italy}

\begin{document}

\maketitle

\section{Abstract}
In this paper I  report the highlights of the talk:   ``Universal properties in galaxies and cored DM profiles'',  given  at: Colloquium  Lectures, Ecole Internationale d'Astrophysique Daniel Chalonge.
The 14th Paris Cosmology Colloquium 2010 ``The Standard Model of the Universe: Theory and Observations''.

\section{Highlights}
The presence of large amounts of unseen matter in  galaxies, distributed  differently from   stars and gas,  is well established   from rotation curves (RCs) which do not show the expected Keplerian fall-off at large radii (Rubin \emph{et al.} 1980),  but  increase, remain flat or start to  gently decrease according to a well organized pattern that involves an invisible mass component becoming progressively more abundant at outer radii  and in  the less luminous galaxies (Persic, Salucci \& Stel 1996).

In Spirals we have the best opportunity to study the  mass distribution:  the   gravitational potentials of   a spherical stellar bulge, a dark  halo, a stellar disk and  a gaseous disk 
give rise to an observed  equilibrium circular velocity 

$$V^2_{tot}(r)=r\frac{d}{dr}\phi_{tot}=V^2_b + V^2_{DM}+V^2_*+V^2_{HI}.$$
The Poisson equation relates the surface (spatial)  densities of these components to the corresponding  gravitational potentials. 
The investigation is not difficult: e.g.  $\Sigma_*(r)$,  the surface stellar  density, is proportional (by the  mass-to-light ratio) to the  observed  
surface brightness: 

$$\Sigma_{*}(r)=\frac{M_{D}}{2 \pi R_{D}^{2}}\: e^{-r/R_{D}}$$ and then $$V_{*}^{2}(r)=\frac{G M_{D}}{2R_{D}} x^{2}B\left(\frac{x}{2}\right),$$
 where $M_D$ is the disk mass, $R_D$ the disk length-scale and $B(x)$ a combination of Bessel functions.  

Dark and luminous matter  in spirals are  coupled:   at any  galactocentric radii $R_n$   measured in terms of disk length-scale $R_n \equiv (n/5)\ R_{opt}$  $(R_{opt}=3.2 R_D)$, there is a   {\it Radial}  Tully-Fisher relation (Yegorova \& Salucci 2007),  i.e. a relation between  the local rotation  velocity  $V(R_n)$ and  the  total galaxy luminosity:  
$M_{band} = a_n \log V_n + b_n$.  Spirals present universal features in their kinematics that correlate  with their  global galactic properties (PSS and Salucci \emph{et al.} 2007).

This led to the discovery,  from  3200 individual RCs, of the ``Universal Rotation Curve'' of Spirals   $V_{URC}(r; L)$ (see PSS and Fig. 1), i.e. a function of   galactocentric  radius $r$, that,  tuned by a global galaxy property (e.g. the luminosity), well reproduces, out to the virial radius (Shankar \emph{et al.} 2006), the RC of any spiral  (Salucci \emph{et al.} 2007).  
$V_{URC}$  is the  observational counterpart to which the  circular  velocity profile emerging in cosmological  simulations must comply (link to 
www.youtube.com/user/dvd5film\#p/a/u/1/YcgafVb-WJI
for a 3-D visualization of the URC).

In the same way of  individual RCs,   it underlies  a mass model  that includes   a Freeman disk  and a  DM halo  with  a Burkert profile 
$$ \rho (r)={\rho_0\, r_0^3 \over (r+r_0)\,(r^2+r_0^2)}.$$
$r_0$ is the core radius and $\rho_0$ the central density,  see Salucci \& Burkert 2000 for details. 
We obtain the   structural parameters $\rho_0$,  $r_0$,  $M_D$ by $\chi^2$ fitting the   URC  and  a number of  individual RCs. As result  a set of  scaling laws  among  local and   global galaxy quantities    emerges (see Fig. 2). 
\begin{figure}[t!]
\centering
\vskip 0.3truecm
\includegraphics[width=6.2truecm]{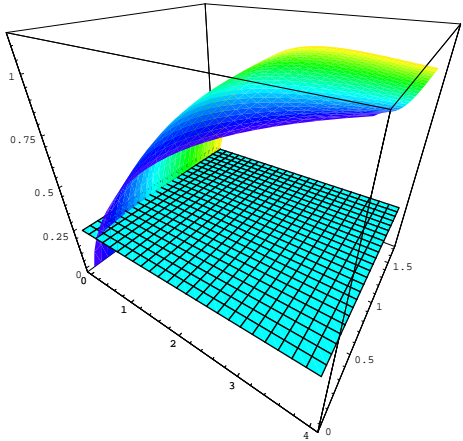}
\vskip -0.1truecm
\caption{The URC. The circular velocity as a function of radius (in units of $R_D$ and  out to $4\  R_D$) and luminosity (halo mass). See Salucci \emph{et al.} 2007 for details and for the URC out to the virial radius.} 
\label{fig:io_urc1}
\end{figure}

These scaling laws indicate (Salucci \emph{et al.} 2007)  that spirals have  an Inner Baryon Dominance region where the stellar disk  dominates  the total gravitational potential,  while the   DM halo emerges  farther out.  At any radii, objects  with lower luminosities have a larger  dark-to-stellar mass ratio. The baryonic fraction in spirals  is always much smaller than the cosmological value $\Omega_b/\Omega_{matter} \simeq 1/6  $, and it ranges between $7\times 10^{-3}$   to   $5\times 10^{-2}$,  suggesting that processes such as  SN  explosions  must have   removed  a very large fraction of the original hydrogen.
Smaller spirals are denser, with their  central density spanning 2 orders of magnitude over the mass sequence of spirals.
\begin{figure}[t!]
\centering
\vskip -0.3cm
\includegraphics[width=8.0truecm]{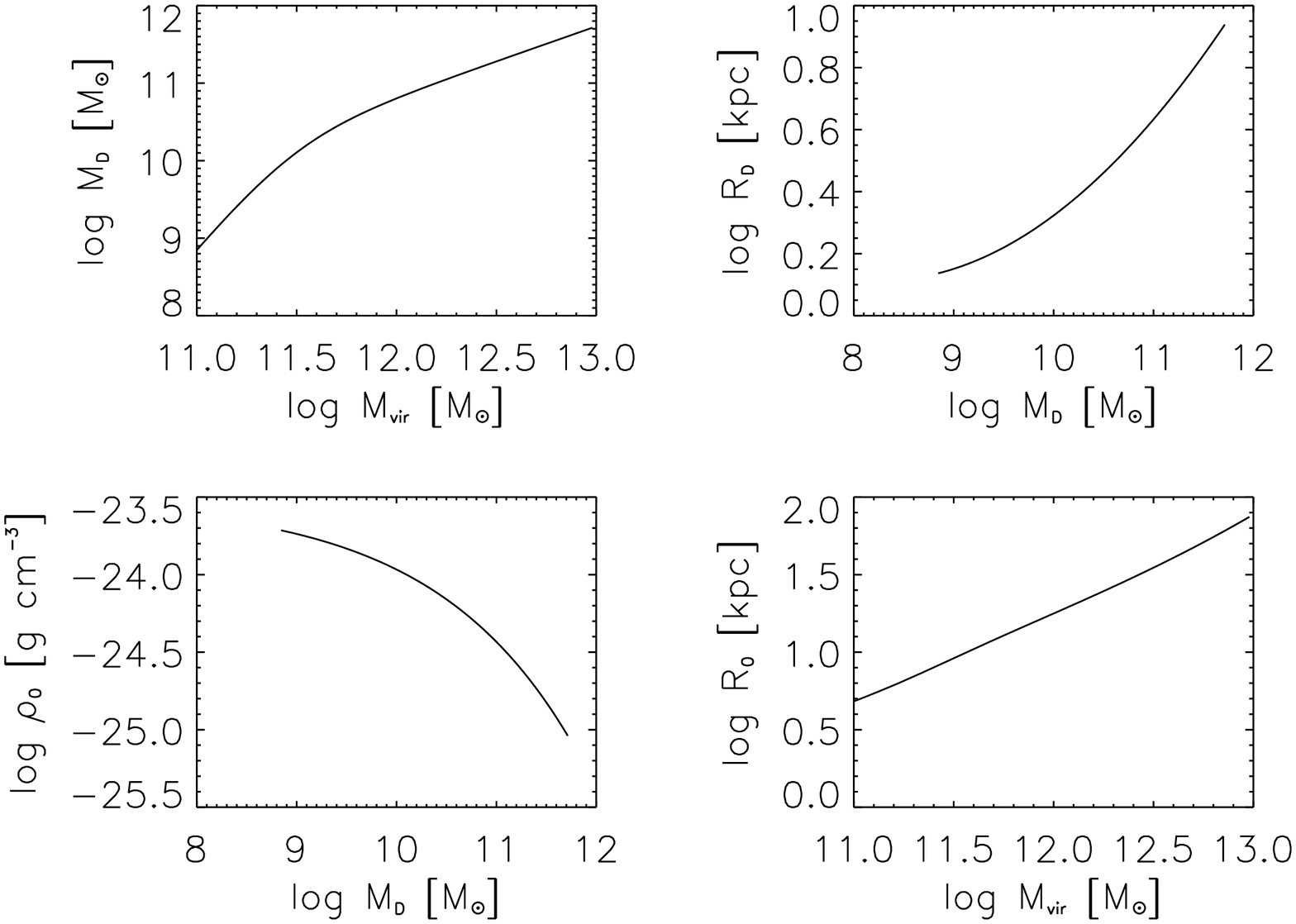}
\vskip -0.6truecm 
\caption{Scaling relations between the structural parameters of the dark and luminous mass distribution in spirals.}
\label{fig:scaling_relations}
\vskip -0.5truecm
\end{figure}

To assume a cored halo profile is obligatory. It is well known that  $\Lambda$CDM  scenario provides a successful picture of the cosmological  structure  formation and that   large N-body  numerical simulations performed in  this scenario  lead to the commonly used NFW  halo cuspy  spatial density profile. However,   a careful analysis  of about 100   high quality,  extended and  free from deviations from axial symmetry  RCs has now ruled out the disk + NFW halo mass model, in favor of cored profiles  (e.g. Gentile \emph{et al.} 2004, 2005, Donato \emph{et al.} 2004, Spano \emph{et al.} 2007, de Blok 2008, Kuzio de Naray \emph{et al.} 2008).

The mass modelling in dSph, LSB and Ellipticals is instead still in its infancy. However, data seem to confirm the pattern shown by  in spirals (Gilmore \emph{et al.} 2005, Walker \emph{et al.} 2010, Nagino \& Matsushita, 2009).

Regarding the structural properties of the DM distribution a  most important   finding is  that   the central surface density  $ \propto \mu_{0D}\equiv  \ r_0 \rho_0$,  where $r_0$ and  $\rho_0$ are  the halo  core radius and central spatial density, is nearly constant and  independent of galaxy luminosity.   Based on the co-added RCs of $\sim 1000$
spiral galaxies, mass models of individual dwarf irregular 
and spiral galaxies of late and early types  with
high-quality RCs and on  galaxy-galaxy weak lensing signals from a sample of spiral and elliptical
galaxies, we find that:  
 $$ \log \mu_{0D} = 2.15 \pm 0.2$$
 in units of $\log$(M$_{\odot}$
pc$^{-2}$).  This constancy  transpasses the family of disk systems and reaches spherical systems.  The  internal kinematics of Local Group dwarf spheroidal galaxies  are
consistent with this picture. Our results are obtained for galactic systems spanning over 14  magnitudes, 
belonging to different Hubble Types, and whose mass profiles have been determined by  several independent 
methods. Very significantly, in  the same objects, the approximate constancy of $\mu_{0D}$ is in sharp contrast to  the
systematical  variations, by several orders of magnitude,  of galaxy properties, including $\rho_0$  and
central  stellar surface density, see Fig. 3.

The evidence that the DM halo central surface density $\rho_0 r_0$
remains constant to within less than a factor of two over fourteen galaxy magnitudes, and across
several Hubble types,  does indicates  that this quantity  is perhaps  
hiding   the  physical nature of the DM. Considering that DM haloes are (almost)  spherical systems it is  surprising that their central surface
density plays  a  role in galaxy structure.   
Moreover it is difficult to  understand this evidence in  an evolutionary scenario as  the product of the process that has turned the primordial cosmological  gas in the stellar galactic structures we  observe today. Such constancy, in fact,   must be achieved in very different  galaxies of different  morphology and mass,   ranging  from
dark matter-dominated  to baryon-dominated objects. In
addition, these galaxies have experienced significantly different
evolutionary histories (e.g. numbers of mergers, significance of
baryon cooling, stellar feedback, etc). 

The best explanation for our findings  relays with the nature itself of the DM, as it seems to indicate recent theoretical  work (de Vega \emph{et al.} 2009, 2010.)  In Warm Dark Matter scenario, 
it is quite possible, for certain values of the particle mass,  to form cored DM virialized  structures (de Vega \emph{et al.} 2010).

The results obtained so far indicate  that the   distribution of matter in galaxies  is a benchmark for  understanding  the dark matter nature   and the  galaxy formation process (at darkmatteringalaxies.selfip.org the reader could  be interested  in  an  initiative that strongly  takes this point of view). In particular, the  universality of certain structural quantities and the dark-luminous coupling  of the  mass distributions  seem to  bear the direct imprint of the Nature of the  DM (Donato \emph{et al.} 2009, Gentile \emph{et al.} 2009).

\begin{figure}[t!]
 \vskip -.0truecm
\psfig{file=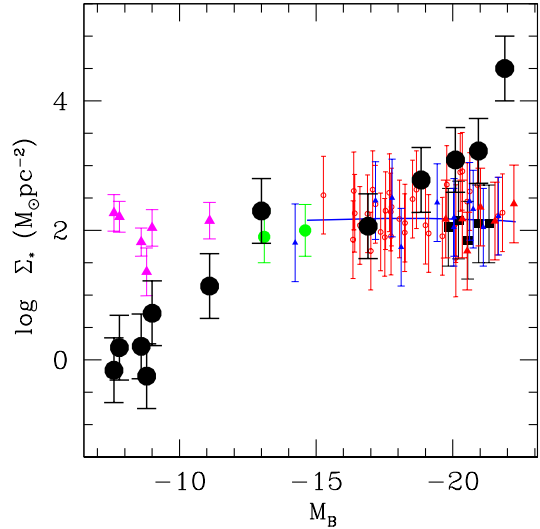,width=0.5\textwidth}   
\vskip -0.8truecm
\caption{Dark matter central surface density in units of $M_\odot$pc$^{-2}$  as a function of
galaxy magnitude,  for different galaxies and Hubble Types. As  a comparison, we also  plot the   values of the same quantity relative to  the stellar component  (big filled circles). }
\end{figure}

\end{document}